\documentclass[conference]{IEEEtran}
\IEEEoverridecommandlockouts
\usepackage{cite}
\usepackage{amsmath,amssymb,amsfonts}
\usepackage{algorithmic}
\usepackage{graphicx}
\usepackage{textcomp}
\usepackage{xcolor}
\usepackage{acronym}
\usepackage{graphicx}
\usepackage{tikz}
\usetikzlibrary{arrows.meta}
\usetikzlibrary{positioning}
\usepackage{amsmath}
\usetikzlibrary{arrows.meta}
\usetikzlibrary{positioning}
\usepackage{amsmath}
\usepackage{microtype}
\usepackage{flushend}

\def\BibTeX{{\rm B\kern-.05em{\sc i\kern-.025em b}\kern-.08em
    T\kern-.1667em\lower.7ex\hbox{E}\kern-.125emX}}

\begin{document}

\title{Improving Radio SLAM via Information Sharing: Collaborative Performance Bounds
\thanks{\textls[-3]{The authors gratefully acknowledge funding from European Union's Horizon Europe Research and Innovation programme under the Marie Sklodowska Curie grant agreement No. 101119643 (MiFuture: ultra-massive MIMO for future cell-free heterogeneous networks, 
https://mifuture.tsc.uc3m.es). The work was also partially supported by the Research Council of Finland under grant \#359095.}
}
}

\author{
\IEEEauthorblockN{
Radovan~Juran$^{\dagger,\star}$,
Ossi~Kaltiokallio$^{\dagger}$,
Julia~Equi$^{\ddagger}$,
Jukka~Talvitie$^{\dagger}$,
Elena~Simona~Lohan$^{\dagger}$,
and Mikko~Valkama$^{\dagger}$
}
\IEEEauthorblockA{$^{\dagger}$Tampere Wireless Research Center, Electrical Engineering Unit, Tampere University, Tampere, Finland}
\IEEEauthorblockA{$^{\star}$Department of Telecommunications, Brno University of Technology, Brno, Czech Republic}
\IEEEauthorblockA{$^{\ddagger}$Ericsson Research, Jorvas, Finland}
}

\newcommand{\N}[0]{\ensuremath{\mathcal{N}}}
\newcommand{\E}[0]{\ensuremath{\mathbb{E}}}

\newcommand{\atan}{ \textrm{atan}  }
\newcommand{\blkdiag}{ \textrm{blkdiag} }
\newcommand{\diag}{ \textrm{diag} }

\renewcommand{\vec}[1]{\ensuremath{{\mathbf{#1}}}}
\newcommand{\vecsymbol}[1]{\ensuremath{\boldsymbol{#1}}}

\newcommand{\vs}[0]{\vec{s}}
\newcommand{\vp}[0]{\vec{p}}
\newcommand{\vm}[0]{\vec{m}}
\newcommand{\vx}[0]{\vec{x}}
\newcommand{\vz}[0]{\vec{z}}
\newcommand{\vu}[0]{\vec{u}}
\newcommand{\vf}[0]{\vec{f}}
\newcommand{\vQ}[0]{\vec{Q}}
\newcommand{\vh}[0]{\vec{h}}
\newcommand{\vR}[0]{\vec{R}}
\newcommand{\vJ}[0]{\vec{J}}
\newcommand{\vy}[0]{\vec{y}}
\newcommand{\vD}[0]{\vec{D}}
\newcommand{\vF}[0]{\vec{F}}
\newcommand{\vH}[0]{\vec{H}}
  \newcommand{\vP}[0]{\vec{P}}

\acrodef{BS}[BS]{base station}
\acrodef{UE}[UE]{user equipment}
\acrodef{TX}[TX]{transmitter}
\acrodef{RX}[RX]{receiver}
\acrodef{TOA}[ToA]{time-of-arrival}
\acrodef{AOA}[AoA]{angle-of-arrival}
\acrodef{AOD}[AoD]{angle-of-departure}
\acrodef{OFDM}[OFDM]{orthogonal frequency-division multiplexing}
\acrodef{LOS}[LoS]{line-of-sight}
\acrodef{NLOS}[NLoS]{non-line-of-sight}
\acrodef{SLAM}[SLAM]{simultaneous localization and mapping}
\acrodef{PCRB}[PCRB]{posterior Cram\'er-Rao bound}
\acrodef{CRB}[CRB]{Cram\'er-Rao bound}
\acrodef{MSE}[MSE]{mean squared error}
\acrodef{FIM}[FIM]{Fisher information matrix}
\acrodef{PDF}[PDF]{probability density function}
\acrodef{PEB}[PEB]{position error bound}
\acrodef{HEB}[HEB]{heading error bound}
\acrodef{CEB}[CEB]{clock error bound}
\acrodef{LEB}[LEB]{landmark error bound}
\acrodef{ISAC}[ISAC]{integrated sensing and communications}

%
%

%

%

\maketitle
\bstctlcite{IEEEexample:BSTcontrol}

\begin{abstract}
This paper studies the fundamental performance limits of radio-based \emph{collaborative simultaneous localization and mapping (C-SLAM)} in dense vehicular environments towards the 6G \ac{ISAC} era. We consider a collaborative estimation task consisting of multiple \acp{UE} with unknown local state variables together with joint estimation of unknown states of the environment landmarks or targets, using cellular radio measurements. Specifically, a \ac{PCRB} is derived for such centralized multi-UE cellular radio SLAM with information sharing, capturing the effects of the number of involved \acp{UE}, their geometric configuration, and their mutually different abilities to measure or observe different landmarks. The analysis reveals how information sharing scales with \ac{UE} density, identifies regimes of diminishing returns, and provides estimator-independent limits for collaborative radio-based environment mapping and localization. The provided numerical results validate the proposed framework in a vehicular system context, while also demonstrating substantial gains against more ordinary single-UE radio SLAM approach.
\end{abstract}


\vspace{-2mm}
\section{Introduction}


\acresetall

The \ac{ISAC} is expected to be one of the main new paradigms and key technical enablers in the future 6G networks~\cite{gonzalez-prelcic2024}, transforming the conventional connectivity-driven radio interface into a joint communication--sensing platform with situational awareness capabilities. The ability to sense the environment and extract situational awareness is of fundamental importance, for example in various vehicular applications, allowing to enhance the location awareness and localization capabilities \cite{HenkVTM23, YuCOMMAG25}. Cellular radio measurements-based \ac{SLAM} is one of the most challenging and tempting prospects of 6G \ac{ISAC}~\cite{sel2025, leitinger2018, wymeersch2018}, and is also the main technical scope of this paper.

In general, \ac{SLAM} has its original roots in robotics research in the 1980s~\cite{durrant-whyte2006}, also sometimes being referred to as concurrent mapping and localization (CML)~\cite{fenwick2002cooperative}, where the sensing entity is traditionally referred to as a robot or an agent. Fundamentally, \ac{SLAM}  is a probabilistic state-estimation problem in which an agent infers both its own position (localization) and the structure of the surrounding environment (mapping) -- via a chosen measurement modality, such as LiDAR or camera. Classical \ac{SLAM}  formulations are typically solved by a single agent without prior knowledge of either its location or the environment~\cite{thrun2006}. However, the estimation process has also been extended to multi-agent settings, where several agents jointly operate in the same unknown area~\cite{fenwick2002cooperative,howard2006,sasaoka2016} while sharing information such as measurements or local state estimates~\cite{chen2023}. Such collaborative SLAM (C-SLAM) approach is, in general, known to expand the coverage of unexplored regions or areas, accelerate joint map construction, and improve the overall estimation accuracy and consistency \cite{al-tawil2024,yuan2025,fan2025}.

\begin{figure}[t!]
    \centering
    \includegraphics[width=0.65\linewidth]{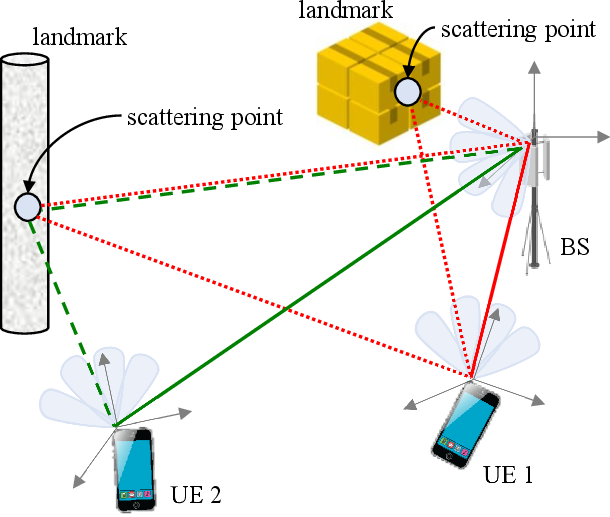}
    \vspace{-2mm}
    \caption{Conceptual illustration of bistatic cellular radio SLAM where individual observable propagation paths of the two involved \acp{UE}, forming the basis for \ac{UE} and landmark location estimation, are highlighted in different colors and line types.}
    \vspace{-4mm}
    \label{fig:scenario}
\end{figure}

The more recent notion of \emph{cellular radio SLAM}, in turn, refers to localizing one or more \acp{UE} while also estimating the coordinates of the environment landmarks or targets with the help of cellular radio measurements \cite{gonzalez-prelcic2024,YuCOMMAG25}. This is illustrated conceptually along Fig.~\ref{fig:scenario}. Compared to robotics \ac{SLAM} where usually only relative state estimation in local frame of reference is pursued, bistatic radio \ac{SLAM} where the network \acp{BS} at known coordinates are utilized as anchors allows extracting \ac{UE} and landmark location information in the global coordinate system \cite{gonzalez-prelcic2024,ge2022mmwave}.


\textls[-5]{One timely and appealing use case of bistatic cellular radio SLAM relates to \emph{future connected vehicles and other vehicular systems} where timely and accurate situational awareness is decisive \cite{gonzalez-prelcic2024,HenkVTM23,YuCOMMAG25}. In general, urban traffic densities span a wide range depending on the city structure and congestion levels. State-of-the-art vehicular use cases consider urban environments with densities reaching up to 12,000 vehicles per square kilometer~\cite{5GAA_CV2X_2025}, represented by heterogeneous traffic composed of both unconnected and connected vehicles~\cite{zhang2025, cheng2025}. Importantly, as the proportion of connected vehicles increases, this density creates a natural advantage and opportunity for \emph{collaborative radio SLAM} -- with more cooperating \acp{UE}, the jointly explored area expands, shared observations accumulate, and the overall estimation uncertainty decreases~\cite{selvaratnam2016}. Although the 12,000~vehicles/km$^{2}$ figure represents the upper extreme of dense metropolitan traffic, the use case specifications in~\cite{5GAA_CV2X_2025} adopt also a reduced density of about 2,000~vehicles/km$^{2}$ as a downscaled but still representative approximation of heavy urban conditions.}

The above provides direct motivation and purpose for this paper and work, in the form of examining and understanding the gains and underlying performance bounds of collaborative radio SLAM. Specifically, in this work, we seek to assess and quantify the benefits of collaboration within dense-traffic vehicular scenarios and derive the corresponding fundamental performance gains and limits using \acp{PCRB}, capturing the impact of cooperating \acp{UE} in a cellular \ac{SLAM} setting independent of any specific parameter estimator scheme. The main scientific novelty and technical contributions of our paper stem from the following three directions or aspects: (\emph{i}) deriving the \ac{PCRB} of collaborative bistatic radio \ac{SLAM} for the first time in the literature, to the best of the authors' knowledge; \emph{(ii)} analyzing the impact of the number of collaborative \acp{UE} on the achievable estimation accuracy in terms of \ac{UE} position error bounds, \ac{UE} heading error bounds, \ac{UE} clock error bounds, as well as landmark position error bounds; and {\emph{(iii)} investigating the collaborative performance bounds for a wide variety of parameter settings including the number of landmarks, magnitudes of process and measurement noise, and different \ac{UE} trajectories.} 
The provided numerical results in the considered vehicular scenario demonstrate major accuracy gains in both \ac{UE} state estimation as well as in landmark state estimation, when compared to ordinary single-UE radio \ac{SLAM}.

The rest of this paper is organized as follows: In Section II, we provide the basic problem formulation in terms of the state variables, and the considered mobility and observation models. The proposed \ac{PCRB} expressions for collaborative radio \ac{SLAM} are derived and provided in Section III. Section IV provides numerical results and their analysis, while the conclusions are drawn in Section V. Finally, selected derivation details are provided in the Appendix.

\section{Problem formulation and models}
\label{sec:problem}
\textls[-5]{In this work, we investigate a two-dimensional bistatic downlink setup where the \ac{BS} acts as the \ac{TX} and the \ac{RX} is at the \ac{UE} side.} The \ac{BS} position and orientation are assumed to be known a priori, whereas the \ac{UE} states are unknown. Altogether, there are $M$ \acp{UE} in the environment and the unknown state of the $m$th \ac{UE} is modeled as $\vs_m = [x_m^\mathrm{UE},\, y_m^\mathrm{UE},\, \alpha_m^\mathrm{UE},\, b_m^\mathrm{UE}]^\top \in \mathbb{R}^4$ while the corresponding joint \ac{UE} state is defined as $\vs = [\vs_1^\top, \, \vs_2^\top, \, \ldots, \, \vs_M^\top]^\top \in \mathbb{R}^{4M}$. The state vector of every \ac{UE}, $m \in \{1,\ldots, M \}$, contains the user position $\vp_m^\mathrm{UE} = [x_m^\mathrm{UE},\, y_m^\mathrm{UE}]^\top$, the heading angle $\alpha_m^\mathrm{UE}$, and the clock offset $b_m^\mathrm{UE}$, which captures the timing mismatch between the unsynchronized clocks of the \ac{BS} and $m$th \ac{UE}. 
\textls[-5]{The \ac{BS} and \acp{UE} are equipped with antenna arrays, and we consider a beam-based \ac{OFDM} transmission framework in which beam sweeping is performed at both the \ac{BS} and the \acp{UE}. The \ac{BS} periodically emits known reference signals, such as the positioning reference signal (PRS), that reach the \acp{UE} through multiple propagation paths. The \acp{UE} then extract the channel parameters associated with each resolvable path. For the $i$th path of a given UE, we denote the estimated \ac{TOA}, \ac{AOD}, and \ac{AOA} by $\vz_{i} = [\tau_{i},\, \phi_{i},\, \theta_{i}]^\top \in \mathbb{R}^3$.  Each \ac{NLOS} path arises from scattering or reflection at passive objects or targets in the environment, and the interaction point of the $i$th path is described by $\vm_n = [x_n, \, y_n]^\top \in \mathbb{R}^2$, which we refer to as a landmark. Altogether, there are $N$ static landmarks, and their joint state defines the environmental map, expressed as $\vm = [\vm_1^\top, \, \vm_2^\top, \, \ldots, \, \vm_N^\top]^\top \in \mathbb{R}^{2N}$. It is important to note that in collaborative \ac{SLAM}, the $M$ \acp{UE} share or try to reconstruct the same map, that is, $\vm$.}

\subsection{Kinematic Model}

\textls[-5]{The vehicle dynamics are modeled through a probabilistic state transition model \cite{durrant-whyte2006}
\begin{equation}\label{eq:transition_density}
    p(\vs_k^{(m)} \mid \vs_{k-1}^{(m)},\, \vu_k^{(m)}) = \N(\vs_k^{(m)}; \vf(\vs_{k-1}^{(m)},\, \vu_k^{(m)}), \vQ_k^{(m)}),
\end{equation}
in which $\vs_k^{(m)}$ and $\vs_{k-1}^{(m)}$ are the $m$th \ac{UE} state at times $k$ and $k-1$, respectively, $\vu_k^{(m)}$ the control input, $\vQ_k^{(m)}$ the process noise covariance, and $\N(\cdot)$ denotes a Gaussian probability distribution. In the model above, the evolution of the state is treated as a Markov process such that the state of the $m$th \ac{UE} at time $k$ is determined solely by the previous state $\vs_{k-1}^{(m)}$ together with the applied control input $\vu_k^{(m)}$, and it does not depend on the other \acp{UE}, the past measurements nor the map. The transition model of the $n$th landmark is $\vm_k^{(n)} = \vm_{k-1}^{(n)}$ since the landmarks are assumed static. In the following, the time index is omitted from the landmark state for brevity.}

\textls[-5]{To describe the vehicle motion, we use the kinematic model for the trajectory of the front wheel of a bicycle subject to rolling motion constraints, that is, assuming zero wheel slip \cite{bailey2006_consistency}. Respectively, the synchronization error between the \ac{BS} and \ac{UE} clocks is assumed to follow a random walk model, which is often used as the stochastic process
to describe evolution of the clock error in radio \ac{SLAM} \cite{ge2022mmwave}. Putting the two models together, the kinematic model for the $m$th \ac{UE} can now be expressed as}
\begin{equation}\label{eq:kinematic_model}
    \vf\left(\vs_{k-1}^{(m)}, \vu_k^{(m)}\right) = \begin{bmatrix}
    x_{k-1}^{(m)} + v_k^{(m)} T_k^{(m)} \cos\left(\alpha_{k-1}^{(m)} + \omega_k^{(m)}\right) \\
    y_{k-1}^{(m)} + v_k^{(m)} T_k^{(m)} \sin\left(\alpha_{k-1}^{(m)} + \omega_k^{(m)}\right) \\
    \alpha_{k-1}^{(m)} + \frac{v_k^{(m)} T_k^{(m)}}{B^{(m)}} \sin\left(\omega_k^{(m)}\right) \\
    b_{k-1}^{(m)}
    \end{bmatrix},
\end{equation}
\textls[-2]{where the control input consists of the speed $v_k^{(m)}$ and steering angle $\omega_k^{(m)}$, $T_k^{(m)}$ denotes the sampling interval, and $B^{(m)}$ is the wheelbase between the front and rear axles of the $m$th vehicle.}

\subsection{Observation Model}
The observation model describes the probability of making an observation $\vz_k$ when the vehicle and landmark states are known. Assuming the measurements are independent and that the noise is zero-mean Gaussian, which are common assumptions in radio \ac{SLAM} \cite{ge2022mmwave,kaltiokallio2021}, the likelihood function is given by
\begin{equation}\label{eq:likelihood}
    p(\vz_k^{(i)} \mid \vs_k^{(m)}, \vm^{(n)} ) = \N(\vz_k^{(i)} \mid \vh(\vs_k^{(m)}, \vm^{(n)}), \vR_k^{(i)}),
\end{equation}
where $\vh(\vs_k^{(m)}, \vm^{(n)})$ is the mean and $\vR_k^{(i)}$ is the measurement noise covariance matrix. It is to be noted that it is reasonable to assume that once the vehicle location and map are defined, observations are conditionally independent given the map and the current vehicle state \cite{durrant-whyte2006}. Furthermore, the data association between the measurements and landmarks is generally unknown and must be estimated together with the unknown landmark and vehicle states.

\textls[-2]{The relationship between the \ac{BS}, $m$th \ac{UE} and $n$th landmark is described by the geometric measurement model, given by \cite{kaltiokallio2025}}
\begin{equation}\label{eq:measurement_model}
    \vh\left(\vs_k^{(m)}, \vm^{(n)}\right) = \begin{bmatrix}
    d_k^{(m,n)}/c + b_k^{(m)} \\
    \atan2\left(-\Delta_{k,y,\phi}^{(m,n)},-\Delta_{k,x,\phi}^{(m,n)}\right) - \alpha_\textrm{BS} \\
    \atan2\left(\Delta_{k,y,\theta}^{(m,n)},\Delta_{k,x,\theta}^{(m,n)}\right) - \alpha_k^{(m)}
    \end{bmatrix},
\end{equation}
in which $\alpha_\textrm{BS}$ is the \ac{BS} orientation and $c$ the speed of light. The model parameters for the \ac{LOS} and \ac{NLOS} paths are defined, respectively, as follows:
\begin{itemize}
    \item \textit{LoS path} -- The signal does not interact with the environment ($\vm^{(n)} = \emptyset$) and the parameters of \eqref{eq:measurement_model} are defined as: $d_k^{(m,n)} = \lVert \vp_\textrm{BS} - \vp_k^{(m)}\rVert$, $\Delta_{k,x,\phi}^{(m,n)} = \Delta_{k,x,\theta}^{(m,n)} = x_\textrm{BS} - x_k^{(m)}$ and $\Delta_{k,y,\phi}^{(m,n)} = \Delta_{k,y,\theta}^{(m,n)} = y_\textrm{BS} - y_k^{(m)}$, in which $\vp_\textrm{BS}$ and $\vp_k^{(m)}$ denote the \ac{BS} and \ac{UE} positions, respectively, and $\lVert \cdot \rVert$ is the Euclidean norm.
    \item \textit{NLoS path} -- The signal interacts once with the environment at location $\vm^{(n)} = [x^{(n)}, \, y^{(n)}]^\top$ as it traverses from the \ac{BS} to the \ac{UE}. The parameters of the model are defined as: $d_k^{(m,n)} = \lVert \vp_\textrm{BS} - \vm^{(n)} \rVert + \lVert \vm^{(n)} - \vp_k^{(m)}\rVert$, $\Delta_{k,x,\phi}^{(m,n)} = x_\textrm{BS} - x^{(n)}$, $\Delta_{k,y,\phi}^{(m,n)} = y_\textrm{BS} - y^{(n)}$, $\Delta_{k,x,\theta}^{(m,n)} = x^{(n)} - x_k^{(m)}$ and $\Delta_{k,y,\theta}^{(m,n)} = y^{(n)} - y_k^{(m)}$.
\end{itemize}

\section{Lower Bounds for Collaborative Radio SLAM}

This section presents a new lower bound for filtering-based collaborative radio \ac{SLAM}. While the \ac{PCRB} is a well-established bound for tracking problems, in general, we are not aware of its application to the collaborative radio \ac{SLAM} problem. The earlier works have only provided the \ac{CRB} for snapshot radio \ac{SLAM} \cite{wymeersch2018} or the \ac{PCRB} \cite{kaltiokallio2021} for filtering-based radio \ac{SLAM} in the single \ac{UE} setting. In the following, the closed-form solution to the \ac{PCRB} for the collaborative radio \ac{SLAM} problem is presented. 

{We note that the bounds derived in this section consider a centralized processing approach in which all the information is available in one place. Such a scenario can be encountered for example in uplink-based measurement scenarios or alternatively also in downlink-based measurements (like here) when the \acp{UE} share all the available information among each other. Deriving bounds for downlink scenarios in which the \acp{UE} only share partial information is also possible, but the bounds presented in this paper are expected to be tighter since all the information is considered.}

\subsection{Posterior Cram\'er--Rao Bound}

In the following, we define the total state as the joint state of the $M$ \acp{UE} and $N$ landmarks $\vx = [\vs^\top, \, \vm^\top]^\top \in \mathbb{R}^{4M+2N}$, and the measurement vector is given by $\vz = [\vz_1^\top, \ldots, \vz_K^\top] \in \mathbb{R}^{3K}$ in which $K$ is the total number of $\vz_i$'s from all \acp{UE}. Furthermore, the data association between $\vz_i$'s and $\vm_n$'s are assumed known since we are interested in the performance bounds. The \ac{CRB} for a time-varying system, commonly referred to as the \ac{PCRB} \cite{tichavsky1995}, states that the \ac{MSE} of an unbiased estimator is always larger than the inverse of the \ac{FIM} in the positive semi-definite sense, mathematically expressed as
\begin{equation}\label{eq:pcrb}
    \E[(\hat{\vx}(\vz) - \vx) (\hat{\vx}(\vz) - \vx)^\top ] \succeq \vJ^{-1},
\end{equation}
where $\vJ$ is the \ac{FIM} and $\hat{\vx}(\vz)$ denotes an estimator of state $\vx$ which is a function of measurement $\vz$.  Let $p(\vx,\vz)$ denote the joint \ac{PDF} of the state and measurements, and let $\Delta_\varrho^\vartheta$ denote the outer product of gradients, $\Delta_\varrho^\vartheta = \nabla_\varrho \nabla_\vartheta^\top$ in which $\nabla_\vartheta^\top = [\tfrac{\partial}{\partial \vartheta_1}, \, \ldots \tfrac{\partial}{\partial \vartheta_L}]$. Now the \ac{FIM} can be defined as
\begin{equation}\label{eq:fim}
    \vJ = \E[-\Delta_\vx^\vx \log p(\vx,\vz)],
\end{equation}
in which the joint \ac{PDF} at time instant $k$ is defined as
\begin{equation}\label{eq:joint_pdf}
    p(\vx_k,\vz_k) = p(\vx_0) \prod_{i=1}^k p(\vz_i \mid \vx_i) \prod_{j=1}^k p(\vx_j \mid \vx_{j-1}),
\end{equation}
in which $p(\vx_0)$ is the prior, $p(\vz_i \mid \vx_i)$ denotes the joint likelihood function and $p(\vx_j \mid \vx_{j-1})$ is the joint transition density. Computing the \ac{FIM} of the joint \ac{PDF} and computing the Schur complement of the block corresponding to all the past states $(\vx_{1:k-1})$, it can be shown that the \ac{FIM} at time step $k$ obeys the recursion \cite{tichavsky1998}
\begin{equation}\label{eq:recursive_fim}
    \vJ_k = \vD_{k-1}^{22} - (\vD_{k-1}^{12})^\top(\vJ_{k-1} + \vD_{k-1}^{11})^{-1} \vD_{k-1}^{12},
\end{equation}
where
\begin{equation}\label{eq:conditional_pdfs}
\begin{split}
\vD_{k-1}^{11} &= \E[-\Delta_{\vx_{k-1}}^{\vx_{k-1}} \log p(\vx_k \mid \vx_{k-1})], \\
\vD_{k-1}^{12} &= \E[-\Delta_{\vx_{k-1}}^{\vx_{k}} \log p(\vx_k \mid \vx_{k-1})], \\
\vD_{k-1}^{22} &= \E[-\Delta_{\vx_{k}}^{\vx_{k}} (\log p(\vx_k \mid \vx_{k-1}) + \log p(\vz_k \mid \vx_{k}) ) ].
\end{split}
\end{equation}
The interpretation of the posterior \ac{FIM} is that $\vD_{k-1}^{22}$ represents the information provided by the data and the prior, and the expression $(\vD_{k-1}^{12})^\top(\vJ_{k-1} + \vD_{k-1}^{11})^{-1} \vD_{k-1}^{12}$ quantifies the loss of information due to propagating the uncertain prior through the probabilistic state transition model.


\subsection{PCRB of Collaborative Radio SLAM}

The conditional \acp{PDF} in \eqref{eq:conditional_pdfs}, for the Gaussian models defined in the previous section, are given by
\begin{equation}\label{eq:conditional_log_pdfs}
\begin{split}
    -\log p(\vx_k \mid \vx_{k-1}) &\propto \lVert \vx_k - \vf(\vx_{k-1},\vu_k) \rVert_{\vQ_k}^2, \\
    -\log p(\vz_k \mid \vx_k) &\propto \lVert \vz_k - \vh(\vx_k) \rVert_{\vR_k}^2,
\end{split}
\end{equation}
in which the notation $\lVert \boldsymbol{\nu} \rVert_{\boldsymbol{\Omega}}^2 = \boldsymbol{\nu}^\top \boldsymbol{\Omega}^{-1} \boldsymbol{\nu}$ is used for brevity. In the above, $\vf(\vx_k,\vu_k)$, $\vh(\vx_k)$, $\vQ_k$ and $\vR_k$ denote the joint kinematic and measurement models, and joint process and measurement noise covariances, respectively. Due to space limitations, we omit the details of the joint models and explicitly give the formulas to compute the \ac{FIM} in Section \ref{sec:fim_computation}. Now, plugging \eqref{eq:conditional_log_pdfs} into \eqref{eq:conditional_pdfs} yields
\begin{equation}\label{eq:conditional_log_gaussians}
\begin{split}
\vD_{k-1}^{11} &= \E[(\nabla_{\vx} \vf(\vx_{k-1},\vu_k)) \vQ_k^{-1} (\nabla_{\vx} \vf(\vx_{k-1},\vu_k)) ^\top], \\
\vD_{k-1}^{12} &= -\E[(\nabla_{\vx} \vf(\vx_{k-1},\vu_k))] \vQ_k^{-1}, \\
\vD_{k-1}^{22} &= \vQ_k^{-1} + \E[(\nabla_{\vx} \vh(\vx_{k})) \vR_k^{-1} (\nabla_{\vx} \vh(\vx_{k}))^\top].
\end{split}
\end{equation}
Computing the expectations above using local linearization and plugging \eqref{eq:conditional_log_gaussians} into the recursive \ac{FIM} formula given in \eqref{eq:recursive_fim}, yields
\begin{multline}
    \vJ_k = \vH_\vx^\top \vR_k^{-1} \vH_\vx + \vQ_k^{-1} \\ 
    - \vQ_k^{-1} \vF_\vx (\vF_\vx^\top \vQ_{k}^{-1} \vF_\vx + \vJ_{k-1})^{-1} \vF_\vx^\top \vQ_k^{-1},
\end{multline}
in which the Jacobians are defined as $\vF_\vx = \nabla_{\vx} \vf(\vx_{k-1},\vu_k)$ and $\vH_\vx = \nabla_{\vx}\vh(\vx_{k})$. Using the Woodbury matrix identity, the above can be simplified to 
\begin{equation}
    \vJ_k = \underset{\vJ_\textrm{prior}}{\underbrace{(\vF_\vx \vJ_{k-1}^{-1} \vF_\vx^\top + \vQ_k)^{-1}}} + \underset{\vJ_\textrm{data}}{\underbrace{\vH_\vx^\top \vR_k^{-1} \vH_\vx}},
\end{equation}
in which $\vJ_\textrm{prior}$ represents the prior information that captures the evolution of the system from time $k-1$ to $k$, and $\vJ_\textrm{data}$ represents the information obtained from the data at time $k$.

After the \ac{FIM} is computed as explained above, the \ac{PEB}, \ac{HEB}, \ac{CEB} and \ac{LEB} for the $m$th \ac{UE} and $n$th landmark can be computed as: $\text{PEB}_k^{(m)} = ( \{ \vJ_k^{-1} \}_{i+1,i+1} + \{ \vJ_k^{-1} \}_{i+2,i+2})^{1/2}$, $\text{HEB}_k^{(m)} = (\{ \vJ_k^{-1} \}_{i+3,i+3})^{1/2}$, $\text{CEB}_k^{(m)} = (\{ \vJ_k^{-1} \}_{i+4,i+4})^{1/2}$, and $\text{LEB}_k^{(m)} = ( \{ \vJ_k^{-1} \}_{j+1,j+1} + \{ \vJ_k^{-1} \}_{j+2,j+2})^{1/2}$, in which $i = 4 (m-1)$ and $j = 4M + 2(n-1)$. The derived lower bounds provide an analytical method for system design and pre-deployment performance evaluation. 

\subsection{Computing the FIM}\label{sec:fim_computation}
\subsubsection{Prior FIM}

Let $\vP_{k-1} = \vJ_{k-1}^{-1}$ denote inverse of the \ac{FIM} at the previous time step, which can be decomposed as
\begin{equation}
    \vP_{k-1} = \begin{bmatrix}
        \vP_{k-1}^{\vs} & \vP_{k-1}^{\vs \vm} \\
        \vP_{k-1}^{\vm \vs} & \vP_{k-1}^{\vm}
    \end{bmatrix},
\end{equation}
where $\vP_{k-1}^{\vs}$ and $\vP_{k-1}^{\vm}$ are the covariances of the \acp{UE} and map, respectively, and $\vP_{k-1}^{\vs \vm}$ is the cross-covariance between the two. The prior \ac{FIM} can be computed by propagating the covariance matrix in time and inverting the predicted covariance back to the information form representation, that is, $\vJ_\textrm{prior} = \vP_{k \mid k-1}^{-1}$. The predicted covariance can be computed by the Chapman-Kolmogorov equation, which yields
\begin{equation}
    \vP_{k \mid k-1} = \begin{bmatrix}
        \vF_{k-1}^\vs \vP_{k-1}^{\vs} (\vF_{k-1}^\vs)^\top + \vQ_k^\vs & \vF_{k-1}^\vs \vP_{k-1}^{\vs \vm} \\
        \vP_{k-1}^{\vm \vs} (\vF_{k-1}^\vs)^\top & \vP_{k-1}^{\vm}
    \end{bmatrix}.
\end{equation}
In the above, $\vQ_k^\vs \in \mathbb{R}^{4M \times 4M}$ is the process noise covariance matrix of the \acp{UE} and $\vF_{k-1}^\vs =\partial \vf(\vs_{k-1}, \vu_k) / \partial \vs_{k-1} \in \mathbb{R}^{4M \times 4M}$ is the Jacobian of the dynamic model evaluated with respect to the joint \ac{UE} state $\vs_{k-1}$. The matrices $\vF_{k-1}^\vs$ and $\vQ_k^\vs$ are block-diagonal, given by
\begin{align}
    \vF_{k-1}^\vs = \blkdiag(\grave{\vF}_{k-1}^{(1)}, \, \ldots, \,  \grave{\vF}_{k-1}^{(M)}), \\
    \vQ_k^\vs = \blkdiag(\acute{\vQ}_k^{(1)}, \, \ldots, \, \acute{\vQ}_k^{(M)}),
\end{align}
in which $\grave{\vF}_{k-1}^{(m)} = \partial \vf(\vs_{k-1}^{(m)}, \vu_k^{(m)}) / \partial \vs_{k-1}^{(m)}$ is the Jacobian evaluated with respect to the $m$th \ac{UE} state $\vs_{k-1}^{(m)}$. The process noise enters the system via noisy control inputs and random clock disturbances which transform the nominal process noise covariance matrix as follows
\begin{equation}\label{eq:process_noise}
    \acute{\vQ}_k^{(m)} = \acute{\vF}_{k}^{(m)} \vQ_k^{(m)} (\acute{\vF}_{k}^{(m)})^\top,
\end{equation}
where $\acute{\vF}_{k-1}^{(m)} = \partial \vf(\vs_{k-1}^{(m)}, \vu_k^{(m)}) / \partial \vu_{k}^{(m)}$. The Jacobians $\grave{\vF}_{k-1}^{(m)}$ and $\acute{\vF}_{k-1}^{(m)}$ are given in the Appendix.

\subsubsection{Data FIM}

Since the measurements are assumed independent, the data \ac{FIM} is additive and $\vJ_\textrm{data}$ can be computed by summing together the \acp{FIM} of the $L$ independent measurements as follows
\begin{equation}
    \vJ_\textrm{data} = \sum_{i=1}^L (\vH_\vx^{(i)})^\top (\vR_k^{(i)})^{-1} \vH_\vx^{(i)}.
\end{equation}
In the above, $\vR_k^{(i)} \in \mathbb{R}^{3 \times 3}$ is the measurement noise covariance matrix of the $i$th measurement (see Eq.  \eqref{eq:likelihood}), and $\vH_\vx^{(i)} =  \partial \vh(\vs_k^{(m)}, \vm^{(n)}) / \partial \vx_k \in \mathbb{R}^{3 \times (4M+2N)}$ is the measurement model Jacobian evaluated with respect to the joint state $\vx$. The Jacobian is non-zero only for the elements involving  $\vs_k^{(m)}$ and $\vm^{(n)}$. As an example, if the $i$th measurement is associated to the first \ac{UE} ($m=1$) and last landmark ($n=N$), the Jacobian is given by
\begin{equation}
    \vH_\vx^{(i)} = \begin{bmatrix}
        \frac{\vh(\vs_k^{(1)}, \vm^{(N)})}{\partial \vs_k^{(1)}} & \mathbf{0}^{3 \times (4(M-1) + 2(N-1))} & \frac{\vh(\vs_k^{(1)}, \vm^{(N)})}{\partial \vm^{(N)}}
    \end{bmatrix}.
\end{equation}
The Jacobians above are given in the Appendix.

\section{Numerical Results}
\label{sec:results}

\subsection{Evaluation Scenario and Setup}

\begin{figure}[!t]
    \centering
    \includegraphics[width=\linewidth]{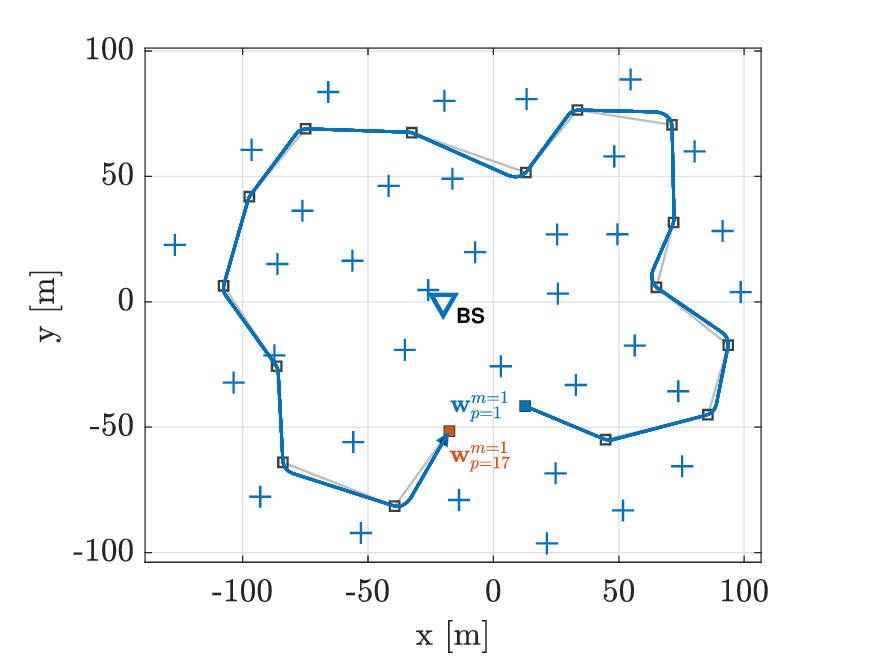}
    \caption[Map illustration.]%
    {The map shows the base station (BS) as a blue triangle
    (\protect\includegraphics[height=0.75em]{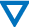}),
    the $N$ landmarks as crosses
    (\protect\includegraphics[height=0.75em]{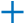}),
    and the path waypoints as squares
    (\protect\includegraphics[height=0.75em]{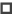}),
    traversed by the UEs starting from $\mathbf{w}^{1}_{1}$
    (\protect\includegraphics[height=0.75em]{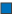})
    to $\mathbf{w}^{1}_{17}$
    (\protect\includegraphics[height=0.75em]{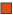}) by default.}
    \label{fig:map-placeholder}
    \vspace{-5mm}
\end{figure}

For the purpose of research reproducibility, the proposed methods are evaluated and analysed in a simulation environment, shown in Fig.~\ref{fig:map-placeholder},
originally developed by Tim Bailey for the robotics community\footnote{Available at: https://openslam-org.github.io/bailey-slam.html}. In this work, the environment has been adapted to account for the geometric characteristics of bistatic radio SLAM. In the experiments, a single \ac{BS} is placed at $[-20,\, 0]^\top \, \text{m}$ and the performance is investigated over varying number of landmarks $N \in \{9,18,35\}$, from which $N=35$ is visualized in Fig.~\ref{fig:map-placeholder}. The \ac{LOS} and single-bounce \ac{NLOS} measurements are modeled using \eqref{eq:likelihood} and corrupted with zero-mean Gaussian noise, with covariance $\vR_k^{(i)} = \diag([1 \, \text{ns}, \, 5 \, \text{deg}, \, 5 \, \text{deg}]^2)$. Measurements are collected every $200 \, \text{ms}$, with the assumption that \ac{LOS} signals are always available, while \ac{NLOS} signals are restricted to landmarks within a $50 \, \text{m}$ radius of the \ac{UE}. 

The unknown states of the $M \in \{1,\ldots,100\}$ UEs evolve according to \eqref{eq:transition_density} with a sampling interval of $T_k = 25 \, \text{ms}$ and a nominal speed of $v_k = 3 \, \text{m/s}$. The turn rate is constrained by $\lvert \omega_k \rvert < 30 \, \text{deg}$, and the covariance of the process noise is defined as $\vQ_k^{(m)} = \diag([1 \, \text{m/s}, \, 10 \, \text{deg}, \, 1 \, \text{ns}]^2)$. The \acp{UE} are initialized using the ground-truth state, while the map is assumed to be unknown at the start, which is standard practice in the robotics community \cite{bailey2006_consistency}. For the considered $200\times200$\,m area, the studied maximum number of UEs ($M=100$) corresponds to a density of $2500$\,vehicles/km$^{2}$, whereas 5GAA specifies a downscaled version of the upper end of urban vehicle density as $2000$\,vehicles/km$^{2}$ \cite{5GAA_CV2X_2025}. Thus, the considered \ac{UE} densities cover realistic vehicle densities from low-load to high-load urban deployments and enable comparative assessment of collaborative vehicular systems over different traffic conditions. Simulations were performed over 100 Monte Carlo trials for each parameter configuration.


The trajectory of the $m$-th UE is defined by a sequence of $p \in \{1,\ldots,17\}$ waypoints $\mathbf{w}^{m}_{p} = [x^{m}_{p},\, y^{m}_{p}]^\top$. For the first UE ($m=1$), the nominal trajectory $\mathbf{w}^{1}_{p}$, shown in Fig.~\ref{fig:map-placeholder}, is always used. The remaining \acp{UE} ($m > 1$) follow randomized variants of the reference waypoints $\mathbf{w}^{1}_{p}$ to consider small realistic deviations from the nominal vehicular path, such as natural variations due to changing lanes or minor drifting. To this end, the $p$-th waypoint of $m$-th UE for $m>1$ is obtained as
\begin{equation}
    \mathbf{w}^{m}_{p} = \mathbf{w}^{1}_{p} + \boldsymbol{\delta}^{m}_{p},
\end{equation}
where $\boldsymbol{\delta}^{m}_{p} = [r^{m}_{p}\cos\theta^{m}_{p},\; r^{m}_{p}\sin\theta^{m}_{p}]^\top$ denotes the random shift vector in which the direction and magnitude of the random shift are described as $\theta^{m}_{p} \sim \mathcal{U}(0,2\pi)$ and $r^{m}_{p} = \sqrt{R^{m}_{p}}$ with $R^{m}_{p} \sim r_{\max}^2 \times \mathcal{U}(0,1)$. The random waypoints $\mathbf{w}^{m}_{p}$ are located uniformly around $\mathbf{w}^{1}_{p}$ on a disk with a radius of $r_{\max} = 2 \, \text{m}$.


\subsection{Results}
\begin{figure}[!t]
    \centering
    \includegraphics[width=\linewidth]{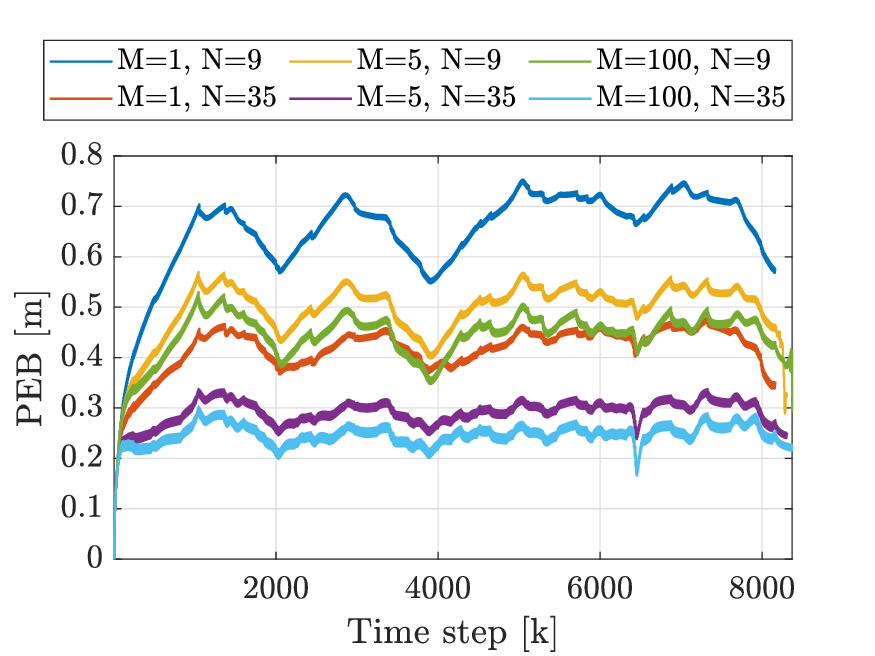}
    \caption{The figure shows the time evolution of the position error bound, PEB, at each epoch $k$, averaged over $100$ Monte Carlo runs, for all considered numbers of UEs $M$ and landmark densities $N$.}
    \label{fig:pcrb_peb_time}
     \vspace{-5mm}
\end{figure}

%
In Fig.~\ref{fig:pcrb_peb_time}, the \ac{PEB} is shown as a function of time step $k$ for different numbers of UEs $M$ and landmarks $N$. The results show that increasing the number of \ac{UE} and landmarks decreases the bound, thus indicating improved positioning performance, while also highlighting evident benefits of \ac{UE} collaboration. The fluctuation of \ac{PEB} in time is mostly due to varying measurement geometry along the \ac{UE} trajectory, and it is generally larger with small number of \acp{UE} and landmarks. Furthermore, although not visible in Fig.~\ref{fig:pcrb_peb_time}, similar trends were observed for the corresponding estimation bounds for the \ac{UE} heading, \ac{UE} clock, and landmark positions.


\begin{figure*}[!t]
\centering
\includegraphics[width=\textwidth]{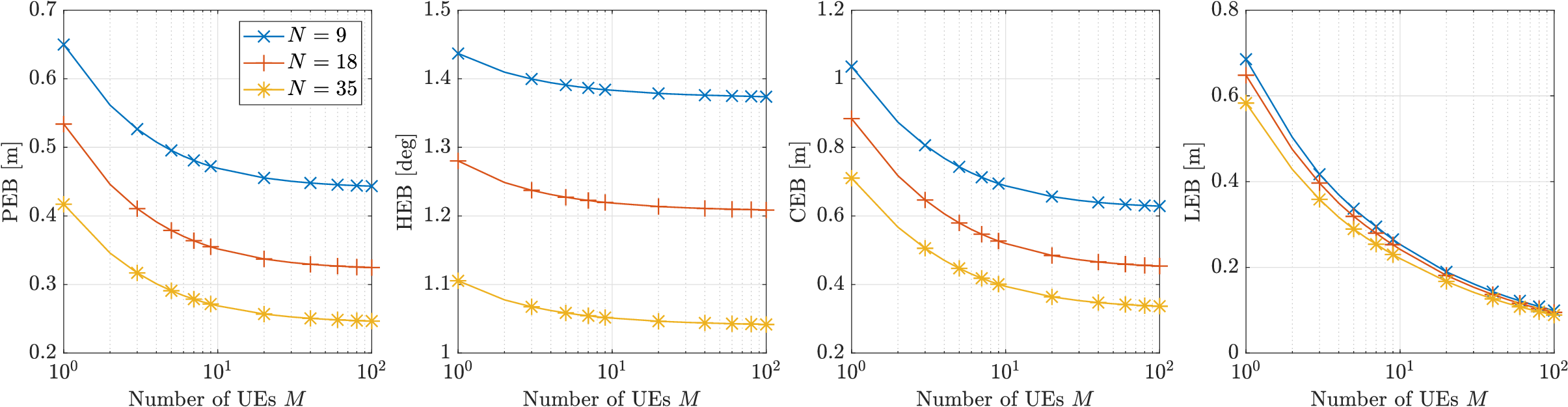}
\caption{The PCRB lower bounds on estimation accuracy averaged over $100$ Monte Carlo simulations per each $M$ of UEs at different landmark density $N$.}
\label{fig:pcrb_per_M}
\vspace{-3mm}
\end{figure*}


The effect of the number of \acp{UE} is further investigated in Fig.~\ref{fig:pcrb_per_M}, which shows the average \ac{PEB}, \ac{HEB}, \ac{CEB} and \ac{LEB} as a function of $M$, while considering different landmark densities. As expected, all bounds decrease as more collaborative \acp{UE} are deployed. When the number of \acp{UE} is small, adding more collaborative \acp{UE} can significantly contribute to the performance bounds until converging at large number of \acp{UE}. The largest improvements in the proposed collaborative performance bounds are observed in the \ac{PEB}, \ac{CEB} and \ac{LEB}, whereas the HEB is affected less. For example, considering $N=18$ landmarks, with $M=100$ UEs the collaborative performance bounds are reduced approximately 40\% for \ac{PEB}, 5\% for \ac{HEB}, 50\% for \ac{CEB}, and 90\% for \ac{LEB} when compared to a single-UE scenario. The \ac{LEB} shows only marginal improvement as $N$ increases. This behavior stems from the fact that the estimation accuracy of landmarks is mainly determined by the measurement geometry and the number of independent observations, which increase with $M$. Consequently, increasing $M$ leads to a more pronounced reduction in \ac{LEB}. Overall, the results demonstrate significant potential of collaborative radio \ac{SLAM} systems with realistic \ac{UE} densities.

\begin{figure}[!ht]
    \centering
    \includegraphics[width=\linewidth]{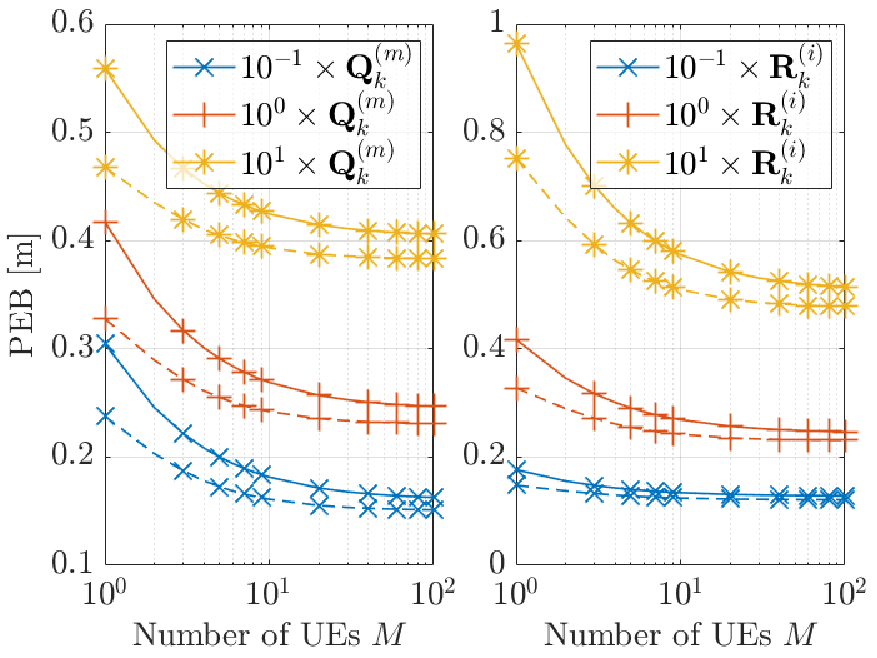}
    \caption{\ac{PEB} as a function of number of \acp{UE} $M$ for different process and measurement noise covariance values, and two different vehicle trajectory scenarios. The counterclockwise trajectory scenario (see Fig.~\ref{fig:map-placeholder}) is illustrated with solid lines and the random trajectory scenario is shown with dashed lines.}
    \label{fig:PEB_vs_QR}
     \vspace{-5mm}
\end{figure}

The effects of vehicle trajectory and the magnitudes of process and measurement noise are further investigated in Fig.~\ref{fig:PEB_vs_QR}, which illustrates the average PEB as a function of $M$. We consider two different trajectory scenarios. In the first scenario, all vehicles follow the counterclockwise trajectory depicted in Fig.~\ref{fig:map-placeholder}, whereas in the second scenario a random permutation of the waypoints is generated for every \ac{UE}. As shown in Fig.~\ref{fig:PEB_vs_QR}, the random trajectory scenario yields lower \ac{PEB} values, mainly because \ac{LEB} converges faster, also having a positive impact on \ac{PEB}. The convergence rate is influenced by two aspects: first,  vehicles explore the entire area faster on average and second, landmarks can be estimated more accurately when \acp{UE} observe the same landmarks from different points of view. 

The magnitudes of process and measurement noise have also a notable impact on \ac{SLAM} performance, and as expected, the performance degrades as the noise increases. Interestingly, \ac{PEB} is doubled when covariance of the process noise is increased by a factor of $100$, whereas the \ac{PEB} increases five fold when the measurement noise is inflated by the same amount. The result thus demonstrates that high performance radio \ac{SLAM} relies more on accurate radio channel measurements rather than precise odometry information of the vehicle. It should also be noted that the importance of collaboration increases as the measurement quality degrades. With very low measurement noise values ($0.1 \times \vR_k^{(i)}$), the collaboration reduced the \ac{PEB} by 27\%, whereas the \ac{PEB} is reduced approximately by 46\% when the measurement noise is extremely high ($10 \times \vR_k^{(i)}$). The opposite is true for the process noise values. Collaboration is more beneficial when the process noise is lower and the impact diminishes as the vehicle dynamics contain more uncertainty.

\section{Conclusion}
Under the assumption of a bistatic cellular downlink scenario with $M$ collaborative \acp{UE} and $N$ static landmarks, we have derived a formulation of {posterior Cram\'er--Rao bounds} (\acp{PCRB}) for radio \ac{SLAM}, which, to the best of our knowledge, have not been reported previously in the literature. Via applying the derived bounds to an example connected vehicles use case, we have further demonstrated that collaborative schemes can significantly enhance the performance, in terms of estimation accuracies of user positions, headings and clock biases as well as the positions of the environment landmarks (i.e., the map). As a concrete example, the collaboration allows decreasing most of the corresponding estimation errors to half when increasing the number of \acp{UE} from one to 100 or more. Thus, based on these newly derived theoretical bounds, collaborative radio \ac{SLAM} emerges as a promising concept for performing localization and mapping within the broad \ac{ISAC} framework in future 6G networks, with particular interest for example in vehicular systems and the related needs for situational awareness. In our future work, the map building and alternative sharing strategies will be investigated and developed to find optimal means of collaboration in practical systems.

\section*{Appendix: Derivation of Jacobians}

Recall that $\vs_{k}^{(m)} = [x_{k}^{(m)},\, y_{k}^{(m)},\, \alpha_{k}^{(m)},\, b_{k}^{(m)}]^\top$ denotes the $m$th \ac{UE} state, $\vu_{k}^{(m)} = [v_{k}^{(m)},\, \omega_{k}^{(m)}]^\top$ the control input of the $m$th \ac{UE}, and $\vm^{(n)} = [x^{(n)},\, y^{(n)}]^\top$ the $n$th landmark. The corresponding Jacobians of the kinematic and measurement models are presented in the following.

\subsubsection{Kinematic Model} The Jacobian of the kinematic model in \eqref{eq:kinematic_model} with respect to the $m$th \ac{UE} state is given by 
\begin{equation}
\frac{\partial \vf(\vs_{k}^{(m)}, \vu_k^{(m)})}{\partial \vs_{k}^{(m)}}
=\begin{bmatrix}
1 & 0 & -v_k T_k \sin(\alpha_{k} + \omega_k) & 0 \\
0 & 1 & v_k T_k \cos(\alpha_{k} + \omega_k) & 0 \\
0 & 0 & 1 & 0 \\
0 & 0 & 0 & 1 \\
\end{bmatrix},
\end{equation}
in which the superscripts have been omitted for brevity. Respectively, the Jacobian with respect to the control input is given by\\ 

\vspace{-2mm}\hspace{1mm}${\partial \vf(\vs_{k}^{(m)}, \vu_k^{(m)})}/{\partial \vu_{k}^{(m)}} $
\begin{equation}
=\begin{bmatrix}
T_k \cos(\alpha_{k} + \omega_k) & -v_k T_k \sin(\alpha_{k} + \omega_k) & 0 \\
T_k \sin(\alpha_{k} + \omega_k) & v_k T_k \cos(\alpha_{k} + \omega_k) & 0 \\
\tfrac{T_k}{B} \sin(\omega_k) & \tfrac{v_k T_k}{B} \cos(\omega_k) & 0  \\
0 & 0 & 1  \\
\end{bmatrix}.
\end{equation}
It is important to note that we have treated $b_k^{(m)}$ as one of the control inputs above for notational convenience. This can be justified by the block diagonal structure of the process covariance matrix which allows expressing $\acute{\vQ}_k^{(m)} = [\acute{\vF}_{k}^{(m)}]_{(:,1:2)} [\vQ_k^{(m)}]_{(1:2,1:2)} ([\acute{\vF}_{k}^{(m)}]_{(:,1:2)})^\top + [\acute{\vF}_{k}^{(m)}]_{(:,3)} [\vQ_k^{(m)}]_{(3,3)} ([\acute{\vF}_{k}^{(m)}]_{(:,3)})^\top$,
in which $\acute{\vF}_{k}^{(m)}={\partial \vf(\vs_{k}^{(m)}, \vu_k^{(m)})}/{\partial \vu_{k}^{(m)}}$, 
conveniently using \eqref{eq:process_noise}.

\subsubsection{Measurement Model (LoS)} The Jacobian of the \ac{LOS} measurement model defined in \eqref{eq:measurement_model} with respect to the $m$th \ac{UE} state is given by 
\begin{equation}
\frac{\partial \vh(\vs_{k}^{(m)}, \vm^{(n)})}{\partial \vs_{k}^{(m)}}
=\begin{bmatrix}
-\Delta_x/d & -\Delta_y/d & 0 & 1 \\
\Delta_y/d^2 & -\Delta_x/d^2 & 0 & 0 \\
\Delta_y/d^2 & -\Delta_x/d^2 & -1 & 0
\end{bmatrix},
\end{equation}
where $\Delta_x = x_\textrm{BS} - x_k^{(m)}$, $\Delta_y = y_\textrm{BS} - y_k^{(m)}$ and $d = \lVert \vp_\textrm{BS} - \vp_k^{(m)} \rVert$. 

\subsubsection{Measurement Model (NLoS)}  Let us define the following auxiliary variables of $\Delta_x^{(1)} = x_\textrm{BS} - x^{(n)}$, $\Delta_y^{(1)} = y_\textrm{BS} - y^{(n)}$, $d_1 = \lVert \vp_\textrm{BS} - \vp^{(n)}\rVert$, $\Delta_x^{(2)} = x^{(n)} - x_k^{(m)}$, $\Delta_y^{(2)} = y^{(n)} - y_k^{(m)}$ and $d_2 = \lVert \vp^{(n)} - \vp_k^{(m)} \rVert$. Now, the Jacobian with respect to the $m$th \ac{UE} state is given by 
\begin{equation}
\frac{\partial \vh(\vs_{k}^{(m)}, \vm^{(n)})}{\partial \vs_{k}^{(m)}}
=\begin{bmatrix}
-\Delta_x^{(2)}/d_2 & -\Delta_y^{(2)}/d_2 & 0 & 1 \\
0 & 0 & 0 & 0 \\
\Delta_y^{(2)}/d_2^2 & -\Delta_x^{(2)}/d_2^2 & -1 & 0
\end{bmatrix},
\end{equation}
and the Jacobian with respect to the $n$th landmark is\\

\vspace{-2mm}\hspace{1mm}${\partial \vh(\vs_{k}^{(m)}, \vm^{(n)})}/{\partial \vm^{(n)}}$
\begin{equation}
=\begin{bmatrix}
\Delta_x^{(2)}/d_2-\Delta_x^{(1)}/d_1 & \Delta_y^{(2)}/d_2-\Delta_y^{(1)}/d_1 \\
\Delta_y^{(1)}/d_1^2 & -\Delta_x^{(1)}/d_1^2  \\
-\Delta_y^{(2)}/d_2^2 & \Delta_x^{(2)}/d_2^2 
\end{bmatrix}.
\end{equation}


\bibliographystyle{IEEEtran}
\bibliography{vtc-spring-2026}

\end{document}